\documentclass[aps,twocolumn,amsmath,amssymb,preprintnumbers,floatfix,prb,superscriptaddress,longbibliography]{revtex4-2}%{revtex4}%

\pdfoutput=1
\usepackage{comment}
\usepackage[version=4]{mhchem}
\usepackage[utf8]{inputenc}
\usepackage{newtxtext}
\usepackage[upint]{newtxmath}
\usepackage{microtype}
\usepackage{textcomp}
\usepackage{dsfont}
\usepackage{eucal}
\usepackage{siunitx}
\usepackage{soul}

\usepackage{enumerate}
\usepackage{amsfonts}
\usepackage{color}
\usepackage{soul}

\usepackage{todonotes}
\presetkeys%
    {todonotes}%
    {inline}{}

\usepackage{graphicx}

\usepackage[colorlinks,allcolors=blue]{hyperref}
\usepackage[capitalize]{cleveref} % Do we want to capitalize?
\usepackage{cleveref}

\definecolor{DarkBlue}{rgb}{0,0,0.80}
\definecolor{DarkRed}{rgb}{0.80,0,0}
\definecolor{Purple}{rgb}{0.55,0,0.55}

\newcommand{\eg}{e.g.\ }

\begin{document}

\title{Voltage-tunable spin supercurrent non-reciprocity reaching 100\% efficiency}
\author{Chi Sun}
\affiliation{Center for Quantum Spintronics, Department of Physics, Norwegian \\ University of Science and Technology, NO-7491 Trondheim, Norway}
\author{Johanne Bratland Tjernshaugen}
\affiliation{Center for Quantum Spintronics, Department of Physics, Norwegian \\ University of Science and Technology, NO-7491 Trondheim, Norway}
\author{Jacob Linder}
\affiliation{Center for Quantum Spintronics, Department of Physics, Norwegian \\ University of Science and Technology, NO-7491 Trondheim, Norway}\date{\today}
\begin{abstract}
The superconducting version of a diode effect has been the subject of extensive research in the last few years. So far, the focus has almost exclusively been on charge transport, but a natural question is whether it is possible to obtain non-reciprocal spin transport without dissipation. Here, we demonstrate that it is possible to generate electrically tunable non-reciprocal spin transport carried by a supercurrent using superconductor/ferromagnet multilayers. The non-reciprocal spin supercurrent reaches an ideal efficiency of 100\%, meaning that the spin-polarization of the critical current is finite in one flow direction, whereas it vanishes in the other direction.
We explain the underlying physics generating this phenomenon. This result provides a way to integrate non-reciprocal supercurrents with spin-polarization, offering new functionality in quantum technologies based on Josephson junctions. 
\end{abstract}

\maketitle

\section{Introduction}
Semiconductor diodes, such as p-n junctions, are one of the most important building blocks of modern electronic circuits, allowing electric current to pass in one direction while being blocked in the opposite direction. Recently, its superconducting analog exhibiting a superconducting diode effect (SDE) \cite{ando_nature_20, nadeem_natrev_23} has garnered extensive attention. This gives nonreciprocity to the supercurrent and further paves the way for the design of dissipationless circuit functionalities. In particular, SDE in the Josephson junction, which is the key element of superconducting circuits, has stimulated growing interest due to its versatile design and construction possibilities. Josephson junction is used in several quantum technologies, including quantum computation based on superconducting logic
circuits, single-photon detectors, switching devices, and quantum bit manipulators. Moreover, quantum computing utilizes non-reciprocal circuit elements to protect qubits from crosstalk and backaction of signals \cite{bardin_jom_21}. The theoretical and experimental realizations include Josephson junctions based on a variety of materials, including a van der Waals heterostructure of $\text{NbSe}_2/\text{Nb}_3\text{Br}_8/\text{NbSe}_2$ \cite{vdW_nature}, a type-II Dirac semimetal $\text{NiTe}_2$ \cite{Pal_2022_NatPhys}, a single magnetic Pb atom \cite{Trahms2023Mar},  magic-angle twisted bilayer graphene \cite{Hu2023Jun}, two-dimensional electron gas \cite{Costa2023Aug}, Rashba spin-orbit coupled quantum dot junctions \cite{debnath_prb_24}, and Andreev molecules \cite{Pillet2023Sep,Hodt2023Nov}. 
%\Cst{Theoretically, the SDE naturally emerges when both space-inversion
%and time-reversal symmetries are broken. Both the sign and magnitude of the SDE can be tuned experimentally via a magnetic field and electric voltage} \cite{sign_reversal_sde_2023,sign_change_2024, TI_SDE_gate, voltage_bias_SDE, three_terminal_sde}.

Theoretically, the SDE naturally emerges in the presence of broken space-inversion
and time-reversal symmetries. An external magnetic field is often used to break time-reversal symmetry to realize SDE. Recently, it has also been experimentally shown that a sign change of the SDE can be realized by tuning the applied magnetic field \cite{sign_reversal_sde_2023,sign_change_2024}. Compared with magnetic field control, the control of SDE by applying voltages is less investigated. In a helical superconductor-based Josephson junction under a biased voltage, it has been predicted that the SDE can be modulated by the voltage, but no sign reversal is achieved \cite{voltage_bias_SDE}. As for the gate voltage control of SDE, a Josephson junction on the surface of a topological insulator offers a sign change of SDE, albeit with a smaller magnitude, by tuning the gate voltage when a strong in-plane magnetic field is applied and the junction length is long \cite{TI_SDE_gate}. Experimentally, a gate-tunable SDE has been studied in a three-terminal Josephson device \cite{three_terminal_sde}, showing that the diode efficiency and polarity can be changed for different electrostatic gating voltages at a given out-of-plane magnetic field.

The research on SDE has focused almost exclusively on charge transport. However, in light
of the importance of spintronics both in terms of practical devices and fundamental research, a natural question is if it is possible to obtain non-reciprocal spin transport without dissipation using superconductors. That is, can one obtain a spin-polarized supercurrent in one direction, whereas the spin-polarization vanishes for current flow in the opposite direction? To the best of our knowledge, this question has only been very recently addressed \cite{spin_SDE} where it was shown that a spin version of SDE is possible, albeit under challenging conditions. (1) It requires a rare $p$-wave superconductor,
which $\text{Sr$_2$Ru}\text{O}_4$ was believed to be \cite{mackenzie_rmp_03}, but recently was experimentally shown \cite{pustogow_nature_19} to not exhibit $p$-wave superconductivity
after all, (2) it requires spin-orbit interactions which are weak in magnitude due to their relativistic origin, resulting
in (3) a limited diode efficiency.

Here, we solve all of the above challenges and also propose a new definition of a non-reciprocal spin supercurrent (NSS), which has a well-defined measurement protocol experimentally. The NSS predicted in this manuscript measures
the spin-polarization of the critical charge supercurrent through the junction, the latter easily tunable via
current-bias, thus circumventing the incapability to exert any experimental control of the spin supercurrent-phase
relation based on present technology. We show that by using (1) conventional superconductors, which are abundant, and (2) no spin-orbit interactions, but rather an electric voltage, it is possible to obtain (3) a NSS efficiency reaching 100\%, meaning that the spin polarization of the critical charge supercurrent is nonzero along one direction while zero in the other direction. In other words, we introduce the concept of a directionally-dependent and electrically controllable spin-polarized supercurrent, and show that it gives rise to an experimentally measurable signature. 
\begin{figure}[t!]
    \centering
    \includegraphics[width = 0.3
    \textwidth]{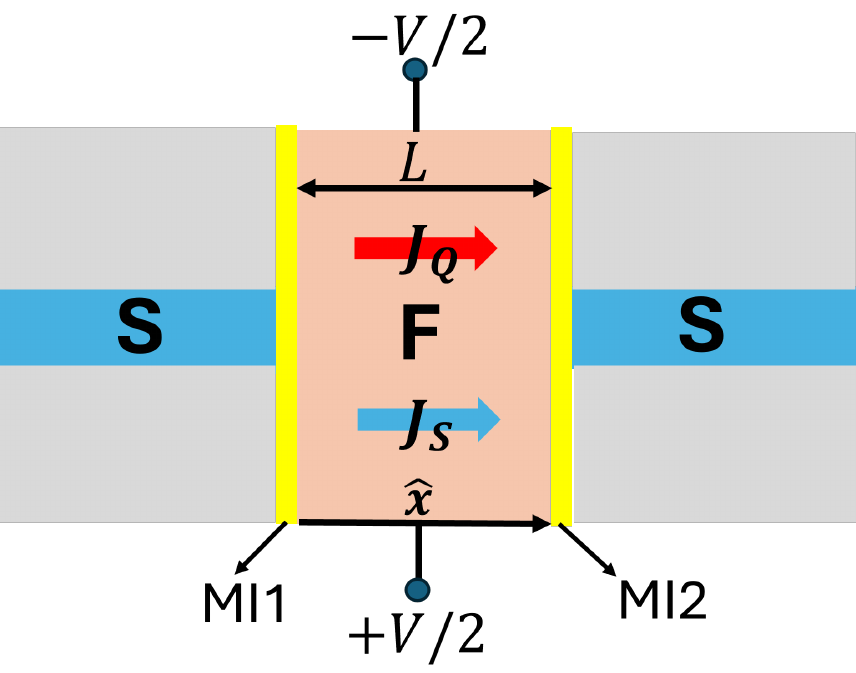}
    \caption{(Color online) A superconductor/ferromagnet/superconductor (S/F/S) Josephson junction on a substrate (gray region) demonstrating a spin-polarized supercurrent transport with 100\% non-reciprocity.  Magnetically misaligned regions are introduced at the superconducting interface through ultrathin magnetic insulators (MIs). The two S electrodes are treated as reservoirs, and have a small width compared to the central F region where the control quasiparticle current flows. The transport of charge supercurrent $J_Q$ and spin supercurrent $\boldsymbol{J}_S$
    can be tuned by the transverse resistive control current which modifies the distribution function of the system, resulting in a voltage-controllable superconducting diode effect (SDE) and non-reciprocal spin supercurrent (NSS), respectively. }
    \label{fig:model}
\end{figure}

\section{Theory} 
We start by providing an intuitive argument for why there can exist a large NSS even for very small deviations from conventional current-phase relationships (CPR) of the Josephson junction. Let $\phi$ be the superconducting phase difference in a Josephson junction. In the simplest case, a conventional charge supercurrent $J_Q(\phi)$ is determined by $\sin\phi$ \cite{golubov_rmp_04}  while a pure spin supercurrent $J_S(\phi)$ is determined by $\cos\phi$. Thus, the phase differences $\phi_\text{max/min} = \pm \pi/2$ which give the maximum and minimum $J_Q$ (i.e. the critical charge current in the left and right direction) are the same values of $\phi$ where $J_S$ vanishes. Consider now a very small deviation away from the ideal $\sin\phi$ and $\cos\phi$ behavior of these currents, i.e. introducing either higher harmonics which skew the CPR relation or an anomalous phase-shift. One can then end up with a situation where the critical charge supercurrent in either direction still occurs for phase differences close to $\pm\pi/2$, but where $J_S(\phi_\text{max}) \to 0$, whereas $J_S(\phi_\text{min}) \neq 0$ has a small, but finite value. The point is then that although the critical charge and spin supercurrents only change by a small amount, the diode efficiency (which is measured by the \textit{relative} difference between the left- and right-going directions) will be very small in the charge current case since both the left- and right-going currents are close to their critical value. In the spin current case, however, the \textit{relative} difference can be very large since the phase differences $|\phi| \simeq \pi/2$ are close to the zeros of $J_S$. This allows for an ideal NSS efficiency of 100\% even for nearly conventional CPR relations. On the other hand, in the case of a CPR relation that strongly deviates from the conventional behavior for the charge- or spin-supercurrent, it is still possible to obtain an efficiency of 100\%. Such deviations can occur in the vicinity of a $0-\pi$ transition \cite{kontos_prl_02, oboznov2006thickness, ryazanov_prl_00, golubov_rmp_04, sellier2004half, braude_prl_07}. Below, we will show examples of both these cases in an S/F multilayer. 

In the mesoscopic S/F/S junction (Fig. \ref{fig:model}), the diffusion of the superconducting condensate into the F can be computed by using the Usadel equation \cite{usadel_prl_70}, which provides a satisfactory description, many times quantitatively, in the typically experimentally
relevant quasiclassical limit. 
The Usadel equation in F reads
\begin{equation}
    D\partial_{\boldsymbol{\hat{x}}}(\hat{g}_F^R \partial_{\boldsymbol{\hat{x}}} \hat{g}_F^R)+i[E\hat{\rho}_3+\hat{M},\hat{g}_F^R]=0,
\label{eq:Usadel_F}
\end{equation}
in which $\hat{g}_F^R$ represents the retarded component of the Green function, $D$ is the diffusion coefficient, $\hat{\rho}_3=\text{diag}(1,1,-1,-1)$, and $E$ represents the quasiparticle energy. The magnetization matrix is given by $\hat{M} = \hat{\boldsymbol{\sigma}}\cdot\boldsymbol{h}_{\text{ex}}$  with $\hat{\boldsymbol{\sigma}}=(\hat{\sigma}_x,\hat{\sigma}_y,\hat{\sigma}_z)$ and $\hat{\sigma}_i=\text{diag}({\sigma}_i,{\sigma}_i^{*})$ where $\boldsymbol{h}_{\text{ex}}$
is the ferromagnetic exchange field. Here $\sigma_i$ represents the Pauli matrices. 
On the other hand, the two S electrodes are considered as reservoirs with the Usadel equation $[E\hat{\rho}_3+\hat{\Delta},\hat{g}_S^R]=0$, in which $\hat{\Delta}=\text{antidiag}(\Delta,-\Delta,\Delta^{*},-\Delta^{*})$. $\Delta$ denotes the superconducting gap and we take it to be $\Delta_0 e^{\pm i\phi/2}$ for the two S layers with $\Delta_0$ being the gap amplitude and $\phi$ the phase difference. The solution of $\hat{g}_S^R$ simply takes the standard Bardeen-Cooper-Schrieffer
(BCS) bulk form. Here the approximations made are precisely the ones that correspond to a typical experimental setting: tunneling interfaces and bulk superconducting electrodes.

To solve for $\hat{g}_F^R$ in Eq. (\ref{eq:Usadel_F}), the boundary conditions at the S/F interfaces are required. Here the spin-active tunneling boundary condition introduced by a magnetic interface (e.g., spin-polarized magnetic insulator) between S and F layers is described as \cite{konstandin_prb_05, Voltage_Ali,spin_active_Ali}
\begin{align}
    2nG_0L\hat{g}_F^R\partial_{\boldsymbol{\hat{x}}}\hat{g}_F^R=G_T[\hat{g}_F^R,F(\hat{g}_S^R)]-iG_\phi[\hat{g}_F^R,\hat{m}],\\
    F(\hat{g})=\hat{g}+\frac{P}{1+\sqrt{1-P^2}}\{\hat{m},\hat{g}\}+\frac{1-\sqrt{1-P^2}}{1+\sqrt{1-P^2}}\hat{m}\hat{g}\hat{m},
\end{align}
in which $n=-1$ for the left (L) S/F interface while $n=+1$ for the right (R) F/S interface. $L$ is the F length and $G_0$ denotes its bulk conductance. Other involved parameters include the interfacial tunneling conductance $G_T$, spin-mixing conductance $G_\phi$ and polarization $P$. The interfacial magnetization is given by $\hat{m}=\hat{\boldsymbol{\sigma}}\cdot\boldsymbol{m}_\text{{L,R}}$. 
The notation $\boldsymbol{m}_\text{L(R)}$ is used for the magnetization of the L (R) interface. Next, we apply the Riccati parametrization \cite{reccati_1995Jul,reccati_arxiv} for the quasiclassical Green function $\hat{g}^R_j$ with $j=S(F)$ to solve $\hat{g}_F^R$ numerically with higher computation efficiency. To model inelastic scattering, a small imaginary part $i\delta$ is added to the
quasiparticle energies $E$ with $\delta/\Delta_0=0.01$.

In our S/F/S system, a voltage is applied transversely across the F which drives the system out of equilibrium by manipulating the occupation states of electrons and holes with the distribution function \cite{SNS_transistor_1998,Heikkila_distribution, pothier1997energy,baselmans1999reversing,Sun2024Dec}
\begin{equation}
    \hat{h}=\frac{1}{2}\left\{\tanh{\left(\frac{E+eV/2}{2T}\right)}+\tanh{\left(\frac{E-eV/2}{2T}\right)}\right\}\hat{\rho}_0,
    \label{eq:distribution}
\end{equation}
where $\hat{\rho}_0=\text{diag}(1,1,1,1)$. This relation is valid near the center of the voltage-biased F under the assumption that the contact area of the superconductor to the ferromagnet is small compared to the width of the ferromagnet \cite{baselmans1999reversing}. Note the above two-step Fermi-Dirac distribution function is  computed from the equation of motion for the Keldysh component in the normal state using voltage-biased reservoirs as boundary conditions \cite{SNS_transistor_1998}. As shown in Appendix \ref{appendix:expressions}, the charge current density $J_Q$ (hereafter referred to simply as current) flowing between the superconducting electrodes is then given by 
\begin{equation}\label{eq: charge current}
    \frac{J_Q}{J_{Q0}} =   \int_{0}^{\infty} \text{Re}\left\{\text{Tr} \left( \hat{\rho}_3\left[ \hat{g}^R_F, \partial_{\boldsymbol{\hat{x}}/\xi_S}\hat{g}^R_F \right]\hat{h}\right) \right\}\frac{dE}{\Delta_0}
\end{equation}
where $J_{Q0}=N_0eD\Delta_0/8\xi_S$, $N_0$ is the density of states at the Fermi level, and $\xi_S$ is the superconducting coherence length.
Before presenting our main results on non-reciprocal spin transport, we first briefly examine the nonreciprocity of the charge transport (supercurrent diode) properties of the system.

\section{Superconducting diode effect} 
The SDE refers to the maximum charge supercurrent in one direction being different from the maximum charge supercurrent in the other direction, which can be mathematically expressed in terms of the charge CPR $J_Q(\phi)$. Here the SDE efficiency is defined as
\begin{equation}
    \eta_Q=\frac{|J_Q(\phi_\text{max})|-|J_Q(\phi_\text{min})|}{|J_Q(\phi_\text{max})|+|J_Q(\phi_\text{min})|},
    \label{eq:eta_e}
\end{equation}
where $\phi \in [-\pi,\pi]$ with $\phi_\text{max}$ ($\phi_\text{min}$) representing the phase difference at which $J_Q(\phi)$ achieves its maximum (minimum). In the absence of SDE, such as for a purely sinusoidal CPR, $\eta_Q=0$. The phase difference $\phi$ is controlled either via an applied charge current through the system or via a magnetic flux in a loop geometry. 

It is known that the diode effects are governed by symmetry properties. The SDE investigated here in general requires that symmetries related to time-reversal and parity (inversion) are broken. In our S/F/S system, the magnetization or exchange field $\boldsymbol{h}_\text{ex}$ in the F forms a spin chirality $\chi$ with the two interfacial magnetizations $\boldsymbol{m}_\text{L,R}$ of the spin-active S/F interfaces according to $\chi=\boldsymbol{h}_\text{ex}\cdot(\boldsymbol{m}_\text{L}\times \boldsymbol{m}_\text{R})$. A nonzero $\chi$ combined with the broken spin-degeneracy satisfies the symmetry requirements for possible diode effect, as $\chi$ is a pseudoscalar that is odd under parity and time-reversal, since the parity operation exchanges $\boldsymbol{m}_\text{L}$ and $\boldsymbol{m}_\text{R}$. To maximize $\chi$, three magnetizations orthogonal to each other are considered, i.e., $\boldsymbol{m}_\text{L}=\hat{\boldsymbol{x}}$, $\boldsymbol{m}_\text{R}=\hat{\boldsymbol{y}}$ and $\boldsymbol{h}_\text{ex}=h_\text{ex}\hat{\boldsymbol{z}}$. Experimentally, perpendicular magnetic anisotropy is known to exist at interfaces such as CoFe/AlO$_x$ \cite{monso2002crossover} and Co/Ni(111) \cite{gottwald2012co}. Moreover, a voltage is applied to change
the occupation of charge-supercurrent-carrying states in F, which importantly provides a purely electrical way of controlling the SDE.

\begin{figure}[b!]
    \centering
    \includegraphics[]{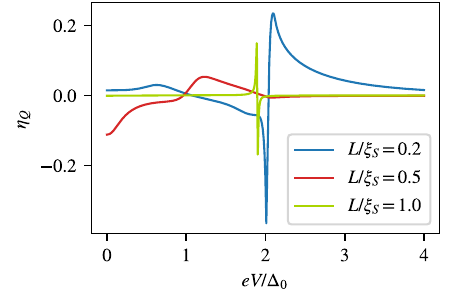}
    \caption{(Color online) SDE efficiency $\eta_Q$ as a function of the applied voltage $V$ for different junction lengths. Here we use 
    $h_\text{ex}/\Delta_0=5$, $P=0.7$, $G_T/G_0=0.3$, 
    $G_{\phi}/G_0 =0.375$ and $T/T_c=0.0176$ with $T_c$ being the critical temperature of the superconductors. The polarization $P\sim 0.7$ is reported in \eg EuS \cite{moodera1988electron} and MgO/Fe/MgO \cite{martinez2018symmetry}. The lengths can readily be rewritten in terms of the ferromagnetic diffusion length $\xi_F$ by $L/\xi_F = \sqrt{h_{\text{ex}}/\Delta_0}\cdot L/\xi_S \sim 2 L/\xi_S$.
     }
    \label{fig:length}
\end{figure}

Figure \ref{fig:length} shows the SDE efficiency $\eta_Q$ as a function of the applied electric voltage for different junction lengths. At equilibrium with $eV/\Delta_0=0$, the SDE strongly depends on the length of the
junction, which is related to the suppression of the first harmonic in the vicinity of a $0-\pi$ transition (see the CPR relations at equilibrium in Appendix \ref{appendix:eta_Q at V=2}).
In the presence of the transverse voltage $V$, it is found that a large magnitude of $\eta_Q$ exceeding $30\%$ can be achieved for $L=0.2\xi_S$ by tuning $V$. 
Around $eV/\Delta_0\sim2$, a sharp jump in $\eta_Q$ occurs for $L=0.2\xi_S$ and $L=\xi_S$ due to a $0-\pi$ transition occurring at this voltage, which gives rise to  a non-standard CPR. This jump can also be understood from the coincidence of an abrupt change in the superconducting Green function and the distribution function elements jumping from zero to unity. Further explanation details are given in Appendix \ref{appendix:eta_Q at V=2}.

\section{Non-reciprocal spin transport}  
 Along with the charge supercurrent $J_Q$, there is spin supercurrent $\boldsymbol{J}_S \sim\boldsymbol{m}_\text{L}\times \boldsymbol{m}_\text{R}$
flowing in the junction which depends on the superconducting phase difference $\phi$. The spin supercurrent exists only inside the Josephson junction. 
Consider the maximized $\chi$ configuration as introduced before, where $\boldsymbol{J}_S=J_S\hat{\boldsymbol{z}}$ is polarized along the exchange field $\boldsymbol{h}_\text{ex}$ in F. Similar to $J_Q$, $J_S$ can be calculated by 
\begin{equation}\label{eq: charge current}
    \frac{J_S}{J_{S0}} =   \int_{0}^{\infty} \text{Re}\left\{\text{Tr} \left( \hat{\rho}_3\hat{\sigma}_z\left[ \hat{g}^R_F, \partial_{\boldsymbol{\hat{x}}/\xi_S}\hat{g}^R_F \right]\hat{h}\right) \right\}\frac{dE}{\Delta_0}
\end{equation}
in which $\hat{\sigma}_z = \text{diag}(1,-1, 1, -1)$
, so that the spin supercurrent is polarized along $\hat{\boldsymbol{z}}$. The derivation details are given in Appendix \ref{appendix:expressions}. The coefficient $J_{S0}$ is obtained by replacing $e$ with $\hbar/2$ in $J_{Q0}$.

Here we introduce and compute the \textit{spin-polarization of the charge supercurrent diode effect}, whose efficiency is defined and characterized via the spin supercurrent evaluated at $\phi_\text{max}$ and $\phi_\text{min}$ as follows:
\begin{equation}
    \eta_S= \frac{|J_S(\phi_\text{max})| - |J_S(\phi_\text{min})|}{|J_S(\phi_\text{max})| + |J_S(\phi_\text{min})|}.
\label{eq:spin_def2}
\end{equation}
This is a quantitative measure of the non-reciprocity of the spin supercurrent (NSS) and we explain later precisely how to measure it. Note here that $\phi_\text{max}$ ($\phi_\text{min}$) have the same definition as introduced before for $J_Q$, corresponding to the phase difference at which $J_Q$ achieves maximum (minimum). This parameter $\eta_S$ measures the difference in spin-polarization between the maximum charge supercurrent flow in each direction in the junction. We underline that this is a different definition than the one very recently proposed in Ref. \cite{spin_SDE}, where the efficiency was defined as the difference in maximum spin-current flowing in each direction. The difference between these definitions is important with respect to the actual measurements of the NSS, and we argue that our definition is experimentally more feasible. The reason for this is that the phases $\phi_\text{max/min}$ are readily applied to the junction since they correspond to the largest charge current-bias that does not generate a voltage drop, i.e. the critical charge supercurrent. Our definition of $\eta_S$ then expresses what the spin-polarization is of the current flowing through the system when it is biased to have maximum charge transport in the left vs. right direction. This is different from identifying the phase differences $\phi$ providing the maximal spin supercurrent flow in each direction, which cannot experimentally be identified by current-biasing the system at the moment, as is normally done for the charge supercurrent.

\begin{figure}[b!]
    \centering
    \includegraphics[]{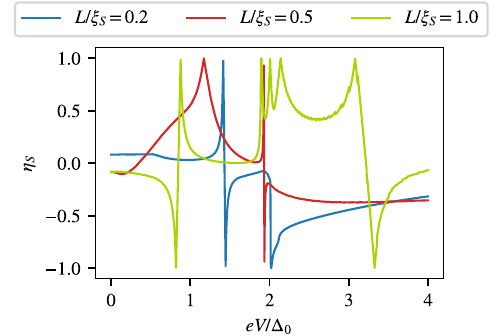}
    \caption{(Color online) NSS efficiency $\eta_S$ as a function of the applied voltage $V$ for different junction lengths. The parameters used are the same as in Fig. \ref{fig:length}. 
    }
    \label{fig:spin_length}
\end{figure}

In Fig. \ref{fig:spin_length}, $\eta_S$ is plotted as a function of the applied voltage $V$ for different junction lengths. It is found that
even the ideal efficiency $|\eta_S|=100\%$ is realizable by tuning the applied voltage.
This ideal value means that the spin polarization of the critical charge supercurrent is zero in one direction and non-zero in the other direction. Moreover, the NSS efficiency factor $\eta_S$ is very large (more than 50\%) over a range of voltages, rather than being confined to a single point in
parameter space. For instance, this happens for $L/\xi_S=0.5$ with $eV$ between approximately $\Delta_0$ and $1.3\Delta_0$ (red curve) or $L/\xi_S=1$ with $eV$ between approximately $1.9\Delta_0$ and $2.4\Delta_0$ (green curve). Within this voltage ranges, one or multiple 100\% points of the efficiency $\eta_S$ are achieved.
We note that $\eta_Q$ and $\eta_S$ can differ greatly in magnitude for a given parameter set, so that a large NSS is not contingent on a large SDE. In addition, only the charge CPR needs to be tuned to obtain a large NSS
(reaching 100\% in some cases), as will be discussed in the following paragraphs. Exerting experimental control over the spin CPR is not required, which is beneficial since it is currently not known how this can be achieved experimentally. 

\begin{figure}
    \centering
    \includegraphics{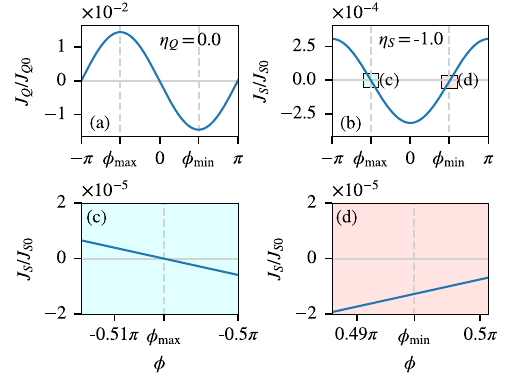}
    \caption{(Color online) (a) Charge and (b) spin CPR at $eV/\Delta_0=3.32$ for $L=\xi_S$. The lower row displays a zoomed-in version of (b) at $\phi_\text{max}$ and $\phi_\text{min}$. The numerical values for $\phi_\text{max}$ and $\phi_\text{min}$ deviate only slightly from $\pm0.5\pi$. However, $|J_S(\phi_\text{max})| \ll |J_S(\phi_\text{min})|$, which causes the NSS efficiency $\eta_S$ to reach the ideal value $-100\%$, even though the charge diode efficiency $\eta_Q=0\%$.
    }
    \label{fig:explain-spin-SDE-2}
\end{figure}

\begin{figure}
    \centering
    \includegraphics{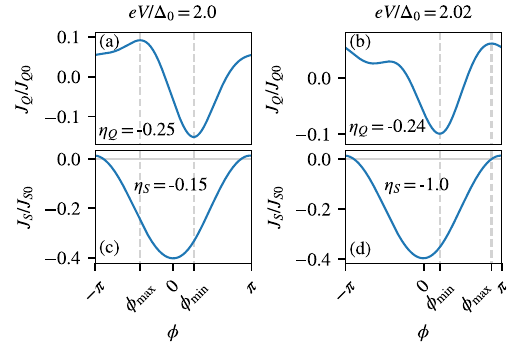}
    \caption{(Color online) Explanation for the observation of an ideal NSS efficiency $\eta_S = -100\%$ for $L=0.2\xi_S$. (a)-(b) show the CPR for $J_Q$. Note the large change in $\phi_\text{max}$ that takes place when increasing the voltage slightly from $eV/\Delta_0=2$ to $eV/\Delta_0=2.02$. (c) demonstrates how the magnitude of the spin supercurrent is comparable at $\phi_\text{max}$ and $\phi_\text{min}$, yielding a small magnitude of $\eta_S$, whereas the large change in $\phi_\text{max}$ causes a large change in the spin supercurrent magnitude in (d). Since $J_S(\phi_\text{max}) \to 0$ in (d), we obtain $\eta_S = -100\%$.}
    \label{fig:explain-spin-SDE-1}
\end{figure}

We now turn to the explanation of the predicted ideal NSS. First, we show how this can be obtained even for a conventional CPR. An example of this occurs for $L=\xi_S$ close to $eV/\Delta_0=3.32$. 
In this case, $\phi_\text{max(min)}$ deviates slightly from $\pm\pi/2$ based on the sinusoidal charge CPR with zero $\eta_Q$ [see Fig. \ref{fig:explain-spin-SDE-2}(a)], but this change is enough to cause a notable change in $\eta_S$ [see Fig. \ref{fig:explain-spin-SDE-2}(b)]. This is because of the $\cos\phi$ behavior of the spin supercurrent which emerges as $eV/\Delta_0$ becomes large, so that both $\phi_\text{max}$ and $\phi_\text{min}$ provide a small magnitude of $J_S$. This allows the NSS efficiency $\eta_S$ to become 100\% for voltages that cause $J_S$ to vanish for either $\phi_\text{max}$ or $\phi_\text{min}$.
This indicates that a charge CPR deviating notably from its standard form $\sin\phi$ is not necessary to realize the ideal NSS: just a conventional $\sin{\phi}$-like behavior of the charge CPR is enough, so long that the system also permits a spin supercurrent which goes like $\cos{\phi}$ and can become zero under the small phase shifts from the ideal $\sin{\phi}$ behavior in $J_Q(\phi)$. The change in the supercurrent magnitude is limited by the small phase shift.

Now, we explore the ideal NSS with
a CPR that strongly deviates from the conventional
behavior. In Fig. \ref{fig:explain-spin-SDE-1}, we plot the charge CPR in the first row to find $\phi_\text{max}$ and $\phi_\text{min}$ for the two voltages $eV/\Delta_0=2$ and $2.02$. These phase differences are used to calculate $\eta_S$ through the spin CPR in the second row. It can be seen that even when the voltage only slightly changes the charge CPR, this is sufficient to cause a large change in $\phi_\text{max}$. When the new $\phi_\text{max}$ is changed to a value where the spin supercurrent $J_S$ vanishes, we get $\eta_S
=-100\%$, indicating that $J_S$ is polarized when the current flows in one direction,
while it is not polarized in the other direction. Note
here the large change of $\phi_\text{max}$ also ensures a large absolute change in the spin supercurrent magnitude. This is beneficial with regard to experimental detection of this effect, which we discuss toward the end of this manuscript.

We have also investigated the robustness of the NSS against different parameter variations (see Appendix \ref{appendix:robustness} for details). We find that the ideal NSS $|\eta_S|=100\%$ remains robust against moderate changes in $h_\text{ex}$, $P$, $G_T$,  $G_\phi$. Moreover, it is reached even for temperatures close to $T_c$. The ideal NSS is also robust against variations in $\chi$, meaning that $\boldsymbol{h}_\text{ex}$, $\boldsymbol{m}_\text{L}$ and $\boldsymbol{m}_\text{R}$ are not fully perpendicular to each other. This robustness towards different parameter variations further facilitates experimental observation of NSS with $\eta_S$ reaching $100\%$ proposed in our work. We also find that the voltage-tunable sign reversal of $\eta_Q$ with a peak magnitude of at least $\sim20\%$ is robust against variations in $h_\text{ex}$, $P$, $G_T$ and $G_\phi$.

%We now explain how the proposed effect can be experimentally detected. The spin supercurrent mediates an RKKY interaction between the magnetic interface layers \cite{slonczewski1989conductance}. This modifies the required strength of an external magnetic field \Chi{$\boldsymbol{H}$} needed to change the magnetization \Chi{$\boldsymbol{M}$}  direction of the layers. Since the spin supercurrent is directionally dependent, the preferred magnetic configuration of the junction can therefore be controlled by both the voltage $V$ providing the transverse current injection and the supercurrent direction itself, \Chi{
%manifesting by measuring $\boldsymbol{M}(\boldsymbol{H})$ influenced by the spin supercurrent-induced torque or effective field. Through a quantitative comparison with the typical current-induced magnetization critical switching current and the coercive field, we assert the spin supercurrent generated in our structure should be large enough to be observable by the proposed experimental setup.} 
%Further details \Chi{regarding the experimental realization} are given in Appendix \ref{Appendix: experiment}. A similar experiment was performed in Ref. \cite{zhu_nmat_17}, albeit without an applied current. There, the difference in the exchange coupling in an F/N/F vs. F/S/F system was detected by measuring the total magnetization at different applied magnetic fields \Chi{$\boldsymbol{M}(\boldsymbol{H})$}, which is what we also propose as the experimental signature for the NSS. 

\section{Proposed Experimental detection setup}

We now explain how the proposed effect can be experimentally detected. The spin supercurrent mediates an RKKY interaction between the magnetic interface layers MI1 with $\boldsymbol{m}_\text{L}=\boldsymbol{\hat{x}}$ and MI2 with $\boldsymbol{m}_\text{R}=\boldsymbol{\hat{y}}$ \cite{slonczewski1989conductance}. Since the spin supercurrent is directionally dependent, the magnetic configuration of the junction can be controlled by both the voltage $V$ providing the transverse current injection and the supercurrent direction itself. Therefore, it is possible to measure the spin current by observing how strong an externally applied magnetic field must be to switch the magnetization $M_x$ along $\boldsymbol{\hat{x}}$. Consider the case where the current is applied in a direction yielding no spin-polarization. At zero applied field, only MI1 with $\boldsymbol{m}_\text{L}=\hat{\boldsymbol{x}}$ will contribute to $M_x$.  If we now apply the external magnetic field $\boldsymbol{H}$ along the $x$-direction, MI2 with $\boldsymbol{m}_\text{L}=\hat{\boldsymbol{y}}$ will eventually align with the applied field once the coercive field is reached, and $M_x$ will increase. Consider instead the case where the current is applied in the opposite direction and there exists a spin supercurrent. The exchange coupling between the MI1 and MI2 is consequently altered. MI2 will now switch its magnetization at a different applied field compared to the scenario where the current was applied in the opposite direction. Thus, the spin supercurrent manifests by measuring $M_x(\boldsymbol{H})$. A similar experiment was performed in Ref. \cite{zhu_nmat_17}, albeit without an applied current, where the difference in the exchange coupling in an F/N/F vs. F/S/F system was detected by measuring the total magnetization at different applied magnetic fields.

We now perform a quantitative analysis on the spin supercurrent-induced torque and the corresponding effective magnetic field. The spin-polarized current-induced 
(anti-damping-like) torque can be described by \cite{brataas_natmat_12,Nernst}
\begin{equation}\label{eq:torque1}
    \boldsymbol{\tau}={-}\frac{\gamma \hbar}{2eM_st_\text{MI2}}\boldsymbol{m}\times(\boldsymbol{m}\times \tilde{\boldsymbol{J}}_S), 
\end{equation}
in which $\gamma$ is the gyromagnetic ratio, $M_s$ is the saturation magnetization, $\boldsymbol{m}$ is the unit vector along the magnetization, and $t_\text{MI2}$ is the thickness of the ferromagnet (MI2 in our case). Note that the spin-polarized current density $\tilde{\boldsymbol{J}}_S$ has the unit of charge current density (A/$\text{m}^2$) since $\hbar/2e$ has been taken outside as the prefactor, i.e., $\tilde{J}_S=\frac{2e}{\hbar}J_S = J_{Q0}\cdot {J}_S/J_{S0}$. Here, $J_S$ is the spin supercurrent density magnitude defined in the manuscript. Therefore, we take the normalized charge supercurrent density constant $J_{Q0}=eN_0D\Delta_0/8\xi_S$ used in the manuscript for the spin supercurrent density $\tilde{J}_S$ above. In our S/MI1/F/MI2/S setup, we consider the magnetization along $\hat{\boldsymbol{x}}$ in MI1, $\hat{\boldsymbol{z}}$ in F and $\hat{\boldsymbol{y}}$ in MI2. To switch the soft magnet MI2, we have $\boldsymbol{m}=\hat{\boldsymbol{y}}$ in Eq. \eqref{eq:torque1}. The spin-polarized current density $\tilde{\boldsymbol{J}}_S$ is polarized along $\hat{\boldsymbol{z}}$ in the F. As a result, we have the torque along $-\hat{\boldsymbol{y}}\times (\hat{\boldsymbol{y}}\times \hat{\boldsymbol{z}})=\hat{\boldsymbol{z}}$.

Take the following typical parameters: 
$e=1.602\times10^{-19}$ C, $N_0=10^{48}$ $\text{J}^{-1}\text{m}^{-3}$, $D={3.8}\times10^{-3}$ $\text{m}^2/\text{s}$, $\Delta_0=1$ meV$=1.602\times10^{-22}$ J, $\xi_S=50$ nm, 
we arrive at $J_{Q0}=2.44\times10^{11}$ A/$\text{m}^2$. In Fig. 5 in the manuscript, we show an example of 100\% NSS efficiency where $J_S=0$ in one direction and $J_S/J_{S0}\approx 0.3$ in the other direction. In this case, $\tilde{J}_S=0.3J_{Q0}={7.32}\times10^{10}$ A/$\text{m}^2$. For current-induced magnetization switching experiments, the typical switching current is around $10^{10}\sim10^{11}$ A/$\text{m}^2$. Based on this, we claim our $\tilde{J}_S$ 
is large enough to make a notable difference on the free magnetization of MI2.

Moreover, the effective field $\boldsymbol{H}_\text{eff}$ that gives the same torque as Eq. \eqref{eq:torque1} is given by
\begin{equation}
    \boldsymbol{H}_\text{eff}={-}\frac{\hbar}{2eM_st_\text{MI2}}\boldsymbol{m}\times \tilde{\boldsymbol{J}}_S.  
\end{equation}
Based on the magnetization configuration described above, it can be easily seen that $\boldsymbol{H}_\text{eff}$ aligns along $-\hat{\boldsymbol{x}}$. 
Consider the additional parameters  $M_s=10^5$ A/m and $t_\text{MI2}=2$ nm. The corresponding effective field magnitude $H_\text{eff}=\frac{\hbar}{2eM_st_{\text{MI2}}}\tilde{J_S}=0.12$ T, which is far above the typical coercive field of a soft magnet (e.g., 0.1 mT). Based on the above comparison with the current-induced magnetization critical switching current and the coercive field, we assert the spin supercurrent generated in our structure should be observable by the experimental setup proposed in our work.

\section{Concluding remarks} 

The superconducting diode effect (SDE)
has been extensively studied, but the spin equivalent is a new concept. We propose a new, experimentally measurable definition for a non-reciprocal spin supercurrent (NSS) that can reach 100\% efficiency 
and which is much more aligned with how experiments on Josephson junctions are typically performed, thus being more practical to work with. To realize the 100\% NSS, three basic factors should be presented in the system: 1) symmetry breaking requirements for the general non-reciprocal effects, 2) magnetic layers to ensure the existence of the spin supercurrent $J_S$, 3) a way to manipulate the charge CPR in the system. We point out that other systems than the specific model considered
in our manuscript fulfilling the above three factors will also allow for the realization of the ideal NSS proposed in our work. The broadness of
the material platforms further facilitates the practicality with regard to
experimental measurements. Our result provides a way to integrate the superconducting diode effect with spin-polarized currents as potential building blocks for quantum computing circuits.

\begin{acknowledgments}
 This work was supported by the Research
Council of Norway through Grant No. 323766 and its Centres
of Excellence funding scheme Grant No. 262633 “QuSpin.” Support from
Sigma2 - the National Infrastructure for High Performance
Computing and Data Storage in Norway, project NN9577K, is acknowledged.   
\end{acknowledgments}

\appendix
\section{Expressions for the charge and spin supercurrents}\label{appendix:expressions}
The charge supercurrent density $J_Q$ flowing between the superconducting electrodes can in general be expressed as energy integrals of the dimensionless charge spectral current  $j_Q$ in F \cite{bergeret_rmp_18},
\begin{equation} \label{eq:JQ_SM}
 J_Q=J_{Q0}\int_0^{\infty}j_Q(E/\Delta_0)dE/\Delta_0,
\end{equation}
where
\begin{equation}
    j_Q(E/\Delta_0)=\text{Re}\left\{\text{Tr}[\hat{\rho}_3(\check{g}_F \partial_{\boldsymbol{\hat{x}}/\xi_S}\check{g}_F)^K]\right\}
\end{equation}
and $J_{Q0}=N_0eD\Delta_0/8\xi_S$. 
Moreover, $\check{g}_F$ is the $8\times8$ Green function matrix in Keldysh space given by 
\begin{equation}
    \check{g}_F=\begin{pmatrix}
    \hat{g}_F^R & \hat{g}_F^K\\
    0& \hat{g}_F^A
    \end{pmatrix}.
\end{equation}
Here, 
$\hat{g}_F^A$ is the advanced component of the Green function which satisfies $\hat{g}_F^A=-\hat{\rho}_3\hat{g}_F^{R\dagger}\hat{\rho}_3$. As for the Keldysh component $\hat{g}_F^K$, it is also related to the non-equilibrium distribution matrix $\hat{h}$ by $\hat{g}^K_F = (\hat{g}_F^R \hat{h} - \hat{h}\hat{g}_F^A)$. Inserting these relations regarding the Keldysh component into Eq. (\ref{eq:JQ_SM}), we obtain 

\begin{equation}\label{eq: charge current}
\frac{J_Q}{ J_{Q0}} = \int_{0}^{\infty} \!\!\!\text{Re}\left\{\text{Tr}\left[ \hat{\rho}_3\left(\hat{g}^R_F(\partial_{\boldsymbol{\hat{x}'}}\hat{g}^R_F)\hat{h}-\hat{h}\hat{g}^A_F\partial_{\boldsymbol{\hat{x}}'}\hat{g}^A_F\right)\right]\right\} \frac{dE}{\Delta_0},
\end{equation}
where $\boldsymbol{\hat{x}}'=\boldsymbol{\hat{x}}/\xi_S$. \
We have set $\partial_{\boldsymbol{\hat{x}}}\hat{h}=0$ since there is no voltage difference between the superconducting electrodes. When the distribution function is diagonal, $\hat{h}= h\hat{\rho}_0$, we can use the relation between the retarded and advanced Green function and the cyclic property of the trace operator to obtain Eq. (5) in the main text,
\begin{equation}
    \frac{J_Q}{J_{Q0}} =   \int_{0}^{\infty} \text{Re}\left\{\text{Tr} \left( \hat{\rho}_3\left[ \hat{g}^R_F, \partial_{\boldsymbol{\hat{x}}/\xi_S}\hat{g}^R_F \right]\hat{h}\right) \right\}\frac{dE}{\Delta_0}.
\end{equation}

Similar to the charge supercurrent density $J_Q$, the spin supercurrent density $J_S$ can be calculated by integrating its corresponding spectral current $j_S$ as 
\begin{equation} \label{eq:JS_SM}
 J_S=J_{S0}\int_0^{\infty}j_S(E/\Delta_0)dE/\Delta_0,
\end{equation}
where
\begin{equation}
    j_S(E/\Delta_0)=\text{Re}\left\{\text{Tr}[\hat{\rho}_3\hat{\sigma}_z(\check{g}_F \partial_{\boldsymbol{\hat{x}}/\xi_S}\check{g}_F)^K]\right\}
\end{equation}
in which $\hat{\sigma}_z=\text{diag}(1,-1,1,-1)$, so that the spin supercurrent is polarized along $\hat{\boldsymbol{z}}$. By applying the relations describing the Keldysh component as used to obtain $J_Q$, we arrive at Eq. (7) in the main text.

\section{Behavior of $\eta_Q$ at $eV/\Delta_0=0$ and $eV/\Delta_0\sim2$}\label{appendix:eta_Q at V=2}

\begin{figure}
    \centering
    \includegraphics[]{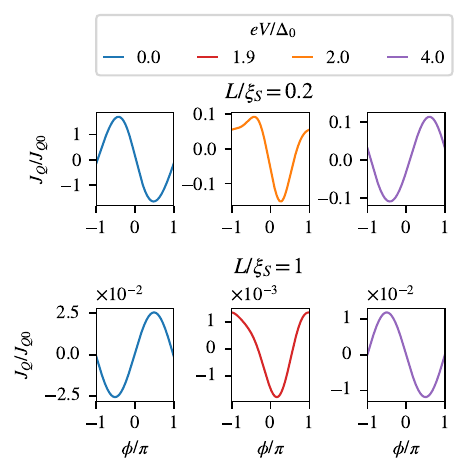}
    \caption{Current-phase relation (CPR) for different voltages when the length of the ferromagnet is $0.2\xi_S$ (upper row) and $\xi_S$ (bottom row). }
    \label{fig:zero-pi transition}
\end{figure}

The jump in the supercurrent diode efficiency $\eta_Q$ close to $eV/\Delta_0=2$ is due to the non-standard current-phase relation that appears in the vicinity of a $0-\pi$ transition of the Josephson junction \cite{kontos_prl_02, oboznov2006thickness, ryazanov_prl_00, golubov_rmp_04, sellier2004half, braude_prl_07}. Figure \ref{fig:zero-pi transition} shows that the junction is a $0-$junction for small voltages and a $\pi$-junction for high voltages when $L=0.2\xi_S$. For the intermediate voltage $eV/\Delta_0=2.0$, the system is close to the transition and the current-phase relation is non-sinusoidal. This causes $\eta_Q$ to be large. The same is seen for $L=\xi_S$. This interpretation is further supported by the non-monotonic length dependence of $\eta_Q$ when $V=0$. When $L=0.2\xi_s$, the junction is then a $0-$junction, and when $L=\xi_S$ it is a $\pi-$junction. We expect $\eta_Q$ to be large for some intermediate length close to the transition point. This is exactly what we see in Fig. 2 in the main text, where $\eta_Q$ is large for $L=0.5\xi_S$ when $V=0$.

\begin{figure}[b!]
    \centering
    \includegraphics[]{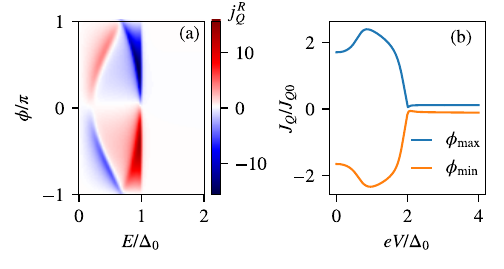}
    \caption{(a) $j_Q^R$ as a function of energy and the phase difference across the junction. (b) The charge current is plotted against the voltage for the phase differences $\phi_{\text{max}}$ and $\phi_{\text{min}}$. The parameters used are the same as in Fig. 2 in the main text ($h_\text{ex}/\Delta_0=5$, $P=0.7$, $G_T/G_0=0.3$, $G_{\phi}/G_0 =0.375$ and $T/T_c=0.0176$) for the length $L/\xi_S=0.2$.} 
    \label{fig:jQR}
\end{figure}

The behavior of $\eta_Q$ at $eV/\Delta_0\sim2$ can also be explained by the coincidence of an abrupt change in the superconducting Green function and the distribution function jumping from 0 to 1. 
The charge spectral current is $j_Q(E/\Delta_0)=j_Q^R(E/\Delta_0)h(E/\Delta_0)$ with

\begin{equation}
     j_Q^R=\text{Re}\left\{\text{Tr} \left( \hat{\rho}_3\left[ \hat{g}^R_F, \partial_{\boldsymbol{\hat{x}}/\xi_S}\hat{g}^R_F \right]\right) \right\}
\end{equation}
We can therefore interpret the distribution function as the weight of $j_Q^R$ in the energy integral defining the charge current.
Figure \ref{fig:jQR}(a) shows the spatial average of $j_Q^R$ as a function of energy and the phase difference across the junction. Here we see that the magnitude of $j_Q^R$ is significantly larger below $E=\Delta_0$ compared to $E>\Delta_0$. This is related to the peak in the anomalous component of $\hat{g}^R_F$ at $E=\Delta_0$, similarly to the bulk superconducting anomalous Green function that is proportional to $1/\sqrt{E^2-\Delta_0^2}$. The nonzero anomalous component shows up in the $\hat{g}^R_F$ due to its proximity to the superconductors. 
The distribution function given by ${h}(E)=\frac{1}{2}\big\{\tanh{[(E+eV/2)/2T]}+\tanh{[(E-eV/2)/2T]}\big\}$ can be simplified to
\begin{equation}
    h(E)=\begin{cases}
        0, 0\le E<eV/2 \\ 1,E>{eV/2}
    \end{cases}
\end{equation}
at zero temperature. At $T>0$ the step is smeared out, but for the sake of the argument, we assume that the step is sharp. When the voltage is small, $eV/\Delta_0<1$, the part of $j_Q^R$ below $E=\Delta_0$ is weighted in the charge current integral, and the current is large. When the voltage is large, $eV/\Delta_0>1$, only the part of $j_Q^R$ above $E=\Delta_0$ where $j_Q^R$ is small is weighted, and the charge current is small. This is shown in Fig. \ref{fig:jQR}(b).
When the voltage is small, $\eta_Q(V)$ will not have any sharp jumps because there are no sharp jumps in $j_Q^R$. Therefore, the current will not change significantly under small variations in $V$. The same applies for high voltages, as seen in Fig. \ref{fig:jQR}(b). However, close to $eV/\Delta_0=2$, small changes in $V$ can cause large changes in the current since there is a jump in $j_Q^R$ at $E\sim\Delta_0$ [see Fig. \ref{fig:jQR}(a)]. Therefore, small changes in the voltage can also cause large changes in the efficiency $\eta_Q.$
  
\begin{figure}[b!]
    \centering
    \includegraphics[width = 0.49\textwidth]{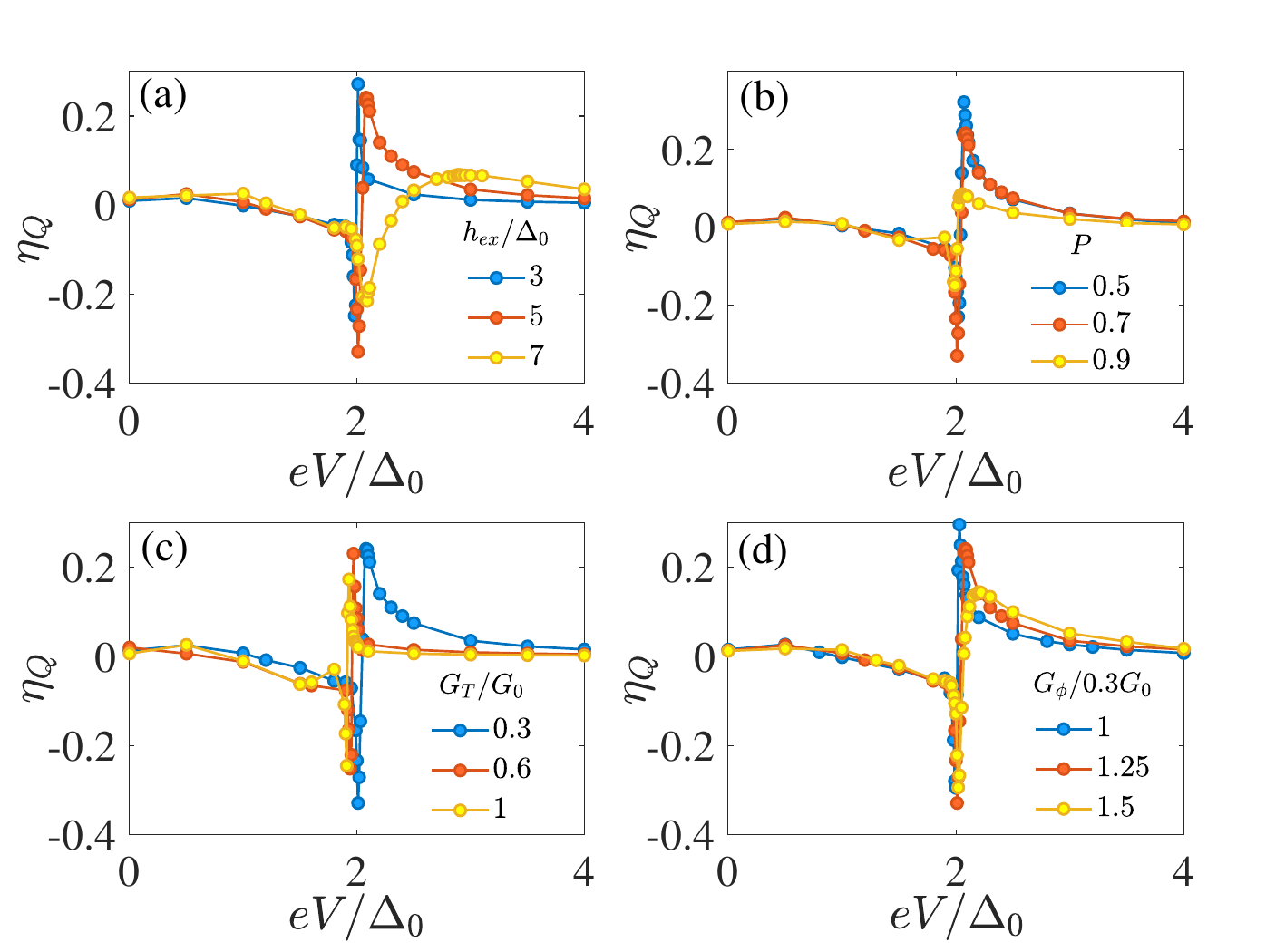}
    \caption{Different parameter dependences of the SDE efficiency $\eta_Q$ for $L=0.2\xi_S$. All other parameters are the same as used in Fig. 2 in the main text except for those indicated in the respective panel. } 
    \label{fig:para1}
\end{figure}

\section{Robustness of the non-reciprocal charge and spin supercurrent against parameter variations}\label{appendix:robustness}

We here demonstrate that both the superconducting diode effect (SDE) and the non-reciprocal spin supercurrent (NSS) in our setup are robust against variations in the different parameters describing the Josephson junction, thus facilitating experimental observation.

\begin{figure}[t!]
    \centering
    \includegraphics[width = 0.5\textwidth]{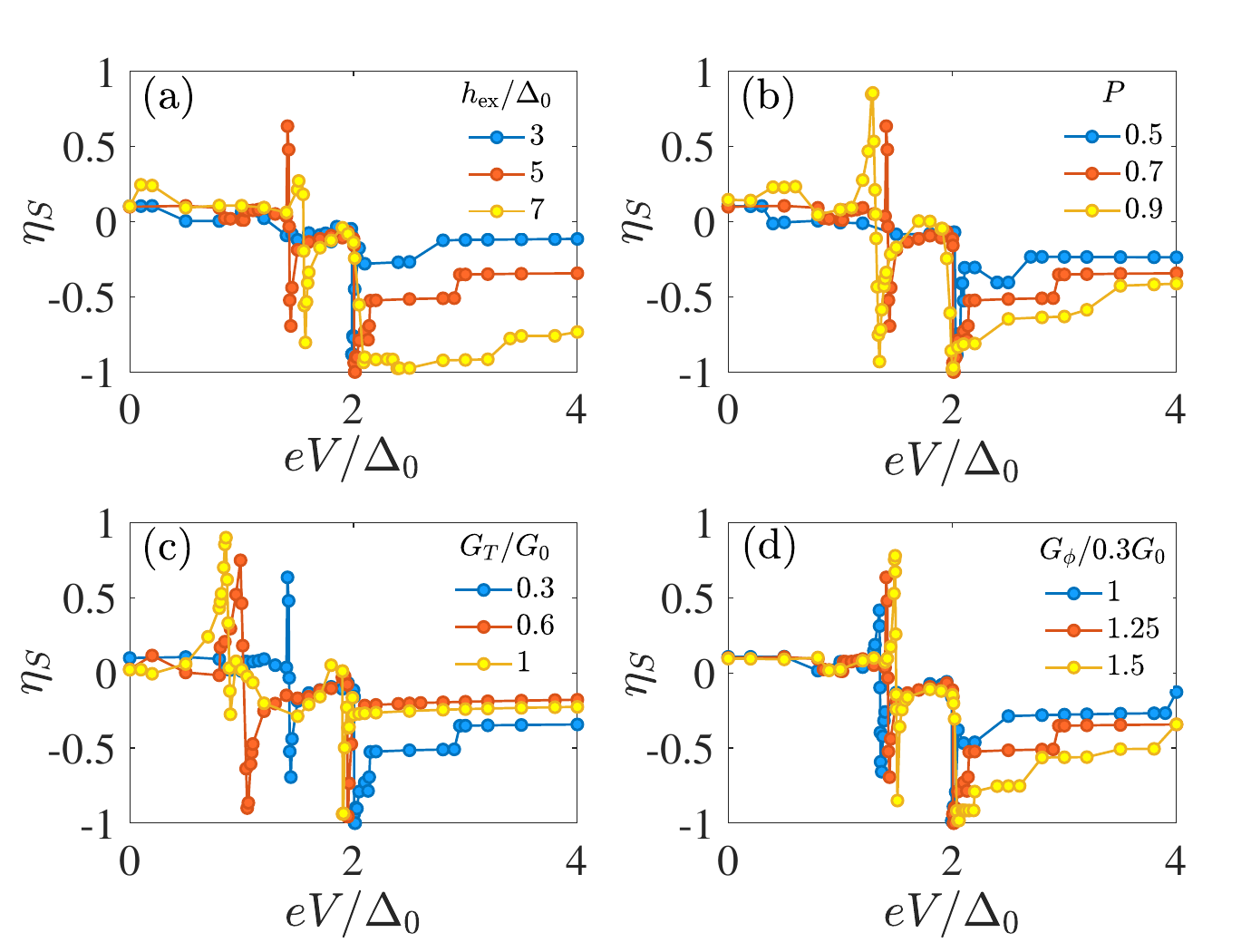}
    \caption{Different parameter dependences of the NSS efficiency $\eta_S$ for $L=0.2\xi_S$. All other parameters are the same as used in Fig. 2 in the main text except for those indicated in the respective panel. } 
    \label{fig:spin_para}
\end{figure}

We first investigate the effects of the exchange field magnitude $h_\text{ex}$ in the F, interfacial polarization $P$, tunneling conductance $G_T$, and mixing conductance $G_\phi$ of the magnetic insulator at the S/F interface. These parameter dependences of the SDE efficiency $\eta_Q$ and NSS efficiency $\eta_S$ for $L=0.2\xi$ are shown in Fig. \ref{fig:para1} and Fig. \ref{fig:spin_para}, respectively. It is found that a larger peak magnitude of $\eta_Q$ generally occurs at values of $h_\text{ex}$ around 5-10 meV (as experimentally realized in \eg PdNi \cite{kontos_prl_02} and Cu$_x$Ni$_{1-x}$ \cite{ryazanov_prl_00}), as well as rather small values of $P$, $G_T$ and $G_\phi$. The voltage-tunable sign reversal of $\eta_Q$ with a peak magnitude of at least $\sim20\%$ nevertheless remains robust against changes in the above parameters.

As for $\eta_S$ in Fig. \ref{fig:spin_para}, it is shown that the moderate variation of $h_\text{ex}$, $P$, $G_T$ and $G_\phi$ changes the voltage where the sharp sign reversal of $\eta_S$ occurs for $eV/\Delta_0<2$. The peak magnitude increases with larger $P$ and $G_\phi$ and therefore provides additional possibilities to reach the ideal $|\eta_S|=1$ when $eV/\Delta_0<2$. On the other hand, the sharp change of $\eta_S$ with the ideal $\eta_S=-1$ achieved at $eV/\Delta_0\sim2$ remains robust against all these parameter variations. In addition, it can be seen that a larger $h_\text{ex}$ widens the range of $V$ where the ideal $\eta_S$ can be maintained for $eV/\Delta_0>2$. We have also confirmed that the ideal $|\eta_S|=1$ is achieved when the $P$ and $G_\phi$ take different values at the two interfaces (not shown here).

\begin{figure}
    \centering
    \includegraphics[]{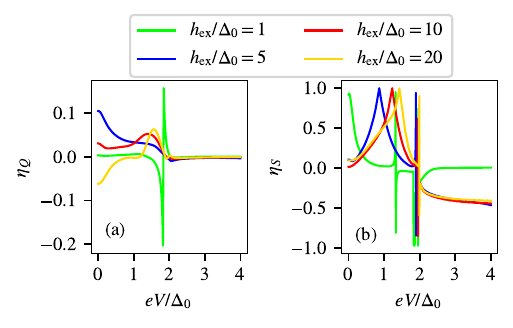}
    \caption{(a) SDE and (b) NSS efficiency at different exchange field strengths $h_\text{ex}$. The length of the ferromagnet is fixed to $L=\xi_F$. The remaining parameters are the same as in Fig. 2 in the main text.}%, \ie $P=0.7$, $G_T/G_0=0.3$, $G_{\phi}/G_0 =0.375$ and $T/T_c=0.0176$.}
    \label{fig:vary-hex}
\end{figure}

Furthermore, the NSS is robust under variations in $h_\text{ex}$ when the length of the ferromagnet is set to $L=\xi_F$. The diffusion length of the ferromagnet is given by $\xi_F=\sqrt{\hbar D/h_\text{ex}}$. The values $h_\text{ex}/\Delta_0=1,5,10,20$ in Fig. \ref{fig:vary-hex} then correspond to $L/\xi_S=1, 0.45, 0.32, 0.22$ respectively, where the superconducting coherence length $\xi_S=\sqrt{\hbar D/\Delta_0}$. The NSS is therefore robust against lengths that are both smaller and longer than the diffusion length of the ferromagnet, and also against different $h_{\text{ex}}$ that determines the relation between $\xi_S$ and $\xi_F$.

\begin{figure}
    \centering
    \includegraphics[width = 0.37\textwidth]{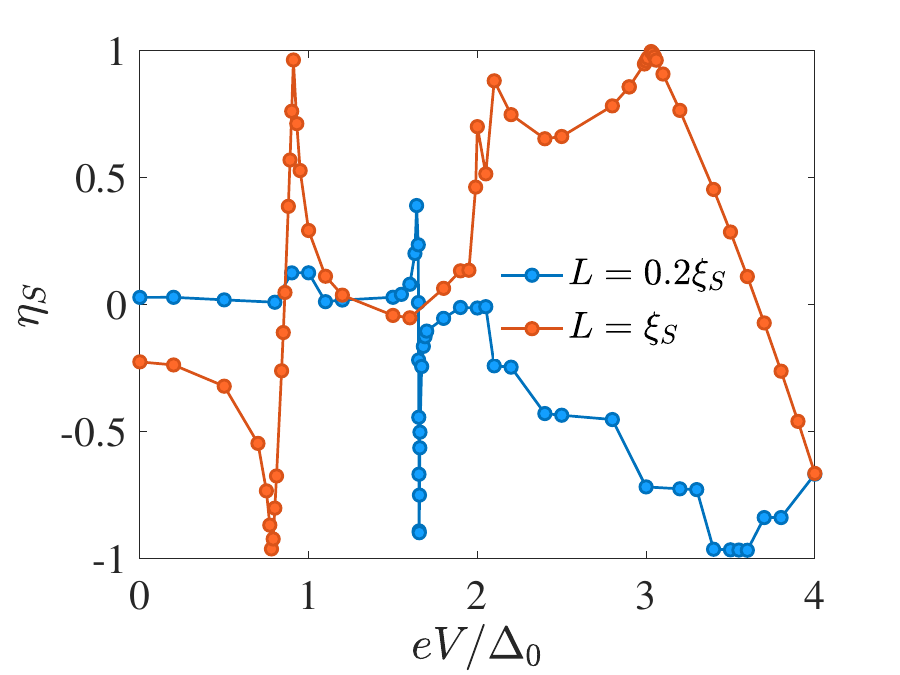}
    \caption{NSS effeciency $\eta_S$ under imperfect $\chi$ [i.e., $\boldsymbol{m}_\text{L}=(\hat{\boldsymbol{x}}+\hat{\boldsymbol{z}})/\sqrt{2}$, $\boldsymbol{m}_\text{R}=(\hat{\boldsymbol{y}}+\hat{\boldsymbol{z}})/\sqrt{2}$ and $\boldsymbol{h}_\text{ex}=h_\text{ex}\hat{\boldsymbol{z}}$]. Note that since the exchange field $\boldsymbol{h}_\text{ex}$ in the ferromagnet is along $\hat{\boldsymbol{z}}$, we are still considering the $z$-component of $\boldsymbol{J}_S \sim\boldsymbol{m}_\text{L}\times \boldsymbol{m}_\text{R}$, which is the only conserved spin supercurrent component in the ferromagnet.
    }
    \label{fig:new_m}
\end{figure}

We next consider the effect of imperfect chirality $\chi$, meaning that the spins are not fully perpendicular to each other, as will likely be the case in an experimental setting. The NSS is shown in Fig. \ref{fig:new_m} to approach an ideal efficiency of 100\% even when the magnetizations are not fully perpendicular to each other, indicating robustness. It is found that the imperfect $\chi$ shifts the sharp change occurring at $eV/\Delta_0\sim2$ for perfect $\chi$ (at small $L$) to larger voltages. Meanwhile, the ideal $\eta_S$ (both positive and negative) becomes achievable for larger $L$ and smaller $V$ with the introduction of imperfect $\chi$, indicating a comparable capability of the imperfect $\chi$ as the maximized $\chi$. As for the $\eta_Q$ (not shown here), its peak magnitude becomes less than $10\%$ for $L=0.2\xi$ while negligible for $L=\xi_S$ with the same imperfect $\chi$ considered in Fig. \ref{fig:new_m}, which again confirms the independence between $\eta_Q$ and $\eta_S$. 

\begin{figure}[]
    \centering
    \includegraphics[]{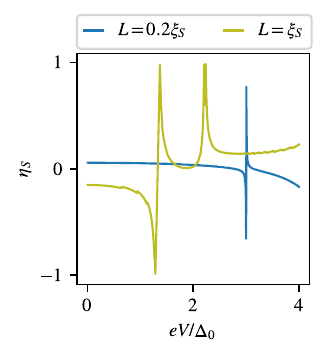}
    \caption{The NSS efficiency $\eta_S$ at $T=0.88T_c$ for two different F lengths. The remaining parameters are the same as in Fig. 2 in the main text.}
    \label{fig:temperature}
\end{figure}

Finally, we consider the NSS efficiency at a higher temperature $T=0.88T_c$. The result is shown in Fig. \ref{fig:temperature}. The magnitude of the SDE efficiency for these parameters is less than 1\% (not shown here), but the NSS efficiency reaches 100\%.

% \clearpage
\bibliography{bib}

% https://arxiv.org/abs/cond-mat/0404383

\end{document}